# TO STUDY THE EFFECTS OF DEGRADATION OF PHONON DISTRIBUTION ON THE HIGH FREQUENCY RESPONSE IN NANO STRUCTURES

*Subir Kumar Sarkar, Souvik Sarkar[1], P. K. Sahu[2], K. Senthil Kumar[3] and S. Rani[4]*

Dept. Electronics and Telecommunication Engineering, Jadavpur University, Kolkata, India   1) RKM Residential College, Narendrapur, India 2)National Institute of Technology, Raurkela, India

3) Dr. M. G. R. College of Engineering, Chenni, India   4) Sapthagiri College of Engineering, Bangalore, India

**ABSTRACT**

Under hot electron condition, the hot carriers phonon excited by ultra-short pulses in polar semiconductors initially loose energy rapidly by emitting longitudinal optic phonons via dominant Frolic coupling. Thus, energy supplied to the carriers by high electric fields goes into phonon generation. As the phonon life time is long enough, phonon distribution is disturbed and a non equilibrium population of LO phonons or hot phonons are produced, leading to their re-absorption by the carriers. In the present work the high frequency performance of GaN nanostructures is studied in the framework of heated drifted Fermi-Dirac distribution function incorporating the relevant scattering mechanisms and the influence of non-equilibrium LO phonons. It is observed that degradation of phonon distribution enhances significantly 3-dB cut off frequency thereby makes the high frequency response flatter reflecting that high frequency response is better if effects of non-equilibrium phonon distributions are included in the calculations.

## 1. INTRODUCTION

Historically, technology has been driving the tools and modes of war even [1] .But nanotechnology is poised to bring an epoch making revolution not only in the weapons, platforms and systems, but also in the very concept of war .Exploitation of nanaotechnology for military has already begun and the synergy among biotechnology, nanotechnology and information technology is being investigated vigorously[2].A review of nanotechnological initiatives launched by various organizations worldwide for weaponisation emphasizes the importance of nanosystems for weapons ,integrated helmet assembly, soldier's uniform and gear, software ,sensors and robotic mules. Applications of nanomaterials for chemical and biological protection, biologically inspired robust software, radar absorption and he application of bucky tubes for antifouling coatings for ships have been examined.

Realization of nanoelectronics materials and devices in parallel with nanofabrication essentially require nanometer-scale evaluation of electrical characteristics as it keeps a direct relevance to the electric or electronic functionality. Electric characteristics are controlled by the system parameters, lattice temperature, external field and frequency of the applied ac electric field [4-6]. All those parameters are related in a complex way so that it is very difficult to predict the best parameters for the most wanted electrical characteristics.

The remarkable advancement in the techniques of crystal of growth such as fine-line lithography, metalorganic of chemical vapor deposition (MOCVD) and molecular beam epitaxy (MBE) has simulated active researches on quasi-low dimensional electronic transport in quantum well wires of width comparable to the electron de-Broglie wavelength [3]. These structures are often referred to as quasi-low dimensional structures (LDS) or nanostructures. They posses radically different properties from those of bulk semiconductors because they quantum mechanically restrict the degrees of freedom of the conduction electrons to two, one or zero. This change in the effective dimensionality offers fascinating changes in electronic, magnetic, optical and vibrational properties. Nanostructures are now recognized as a promising basis for the study of the physics of LDS and their future technological applications. Researches on the physics of LDS continue to be both challenging and exciting, as novel structures with different materials having different properties are developed [4].

In nanostructures, hot electron conditions are developed if the applied electric field is sufficiently high so as to cause a pronounced deviation from Ohm's law. At these is fields, drift energy of electrons may be compared with the thermal energy at the lattice temperature. The average electron energy is also much higher than that in the thermal equilibrium with the lattice. Under such high electric fields the carrier mobility exhibits a complicated dependence on the applied electric field. Under hot electron condition, the hot carriers, photon excited by ultra short pulses in polar semiconductors, initially loose energy rapidly by longitudinal optic phonons via dominant frolic coupling. Thus the energy supplied to the carriers by the electric field goes into phonon generation. As the phonon life time is long enough, phonon distribution is disturbed and a non equilibrium population of LO phonons or hot phonons are produced, leading to their reabsorption by the carriers. The measurement of energy loss rate in low dimensional nanostructures and other analysis show that the non equilibrium LO phonons play an important role in the slowing down the carrier cooling rate. In the present work the high frequency performance of GaN nanostructure is studied in heated drifted Fermi-Dirac distribution function incorporating the relevant scattering mechanisms and the influence of non equilibrium LO phonons. The effect of large q-LO phonon and electron LO phonon interaction at low $qL_z$ ($L_{Z\ is}$ channel length) are also studied.

## 2. MODEL

Consider a semiconductor quantum structures (also called nanostructures) of GaN. For the carrier concentration ($n_{2D}$), channel length ($L_z$), and the lattice temperature used here, the separation between the lowest and thee next higher sub-band is at least four-times the average carrier energy. The electrons are thus assumed to populate only the lowest sub-band of the nanostrctures in an infinite barrier height approximation.

Carrier scattering via ionized impurities, poles optic and deformation potential acoustic phonons are considered here, in the calculations, using a heated-drifted Fermi-Dirac distribution function. Now:

$$f(k) = f_0(E) + (hp_d/m^*)k(-\partial f_0/\partial E)\cos\theta$$
..... (1)

where $f_0(E)$ is the equilibrium Fermi-Dirac distribution function for the carriers characterized by an electron temperature $T_c$, $m^*$ the electron effective mass, $h$ the Plank's constant distant divided by $2\pi$, $\theta$ the angle between the 2D wave, vector k and the applied electric field F, and $p_d$ the drift crystal momentum.

Screened scattering rates are considered, except for LO phonons, where carrier screening is insignificant in thee present condition. A small electric field of magnitude $F_1$ and angular frequency $\omega$ super-imposed on a moderate dc bias field $F_0$ is assumed to act parallel to the hetero junction interfaces. The net field is thus given by:

$$F = F_0 + F_1 \sin\omega t. \quad \ldots (2)$$

In the present of the electron field, the electron temperature $T_c$, the drift momentum $p_d$ and the phonon distribution $N_Q$ will have similar components with the alternating ones, generally differing in phase. Hence, one can write:

$$T_c = T_0 + T_{1r}\sin\omega t + T_{1i}\cos\omega t. \quad \ldots (3)$$
$$p = p_0 + p_{1r}\sin\omega t + p_{1i}\cos\omega t. \quad \ldots (4)$$
and
$$N_Q = N_{Q0} + N_{Qr}\sin\omega t + N_{Qi}\cos\omega t \quad \ldots (5)$$

The phonon generation rate for the 2D system is calculated here, along the same lines as for the bulk material, including the degeneracy of the distribution function [5]. The LO phonons decay towards the thermal equilibrium distribution $N_Q$ with a rate characterized by the phonon life-time. In equation (5), $N_{Q0}$ represent the contribution of the equilibrium distribution $f_0(E)$ term, $N_{Q1} = N_{Qr} + j N_{Qi}$ represents the drift term. The phonon drift and, therefore, the phonon drag are independent of carrier concentration at low carrier concentration. But, at higher carrier concentration (as in the present case), the phonon drag contribution decreases. It is deduced that, scattering of the phonon by the carriers has become sufficiently frequency at these concentrations, to decrease the phonon drift. In high electron fields, there will be an increase resulting from the departure from thermal equilibrium of the $f_0(E)$ or random terms in the carrier distribution. It is already mentioned that the drag or drift term in the disturbed phonon distribution is due to the perturbation term of the heated-drifted Fermi-Dirac distribution function. The exact from of the band structure and scattering processes. The second term of thee distribution function contains two-dimensional wave-vector (k), which is expressed as:

$$K = (e^{ik.r}/\sqrt{A})(\sqrt{2/L_z})\sin(\pi z/L_z) \quad \ldots (6)$$

where, A is the layer area, r is a two-dimensional position vector in the (x,y) plane, k is the 2D wave vector in the (x-y) plane and $L_z$ is the thickness of the quantum well. This supports that, $N_{Qi}$ may be negative. Further, $N_1 = N_{Qr} + j N_{Qi}$. This indicates that, $N_{Qi}$ is providing only some mathematical details, as in communication there arises a term like, negative frequency or stability criteria in control system. Moreover, $N_{Qo}$ is much higher than $N_{Qi}$, so, in on case the phonon occupation negative.

The net time rate of increase of $N_Q$ is given by:

$$(\delta N_Q/\delta t) = (\delta N_Q/\delta t)_{gen} + (\delta N_Q/\delta t)_{decay} \ldots\ldots (7)$$

where the two terms on the right hand side denote the phonon generation and decay rates, reactively.

Also:

$$(\delta N_Q/\delta t)_{gen} = A \int [[(N_Q+1)F^2(q_zL_z)f_0(E+h\omega_0)\{1-f_0(E)\}] / [q(q+q_z^2)\sin\theta_0] - [N_Q F^2(q_zL_z)f_0(E)\{1-f_0(E)\}] / q(q^2+q_z^2)\sin\theta_0]]\,dk \quad \ldots\ldots\ldots(8)$$

where, $A = [[e^2\omega m^*(K_s - K_\infty)]/\pi i^2 L_z K_s K_\infty]$

The first term of integrand Eq. (6) arises, when a carrier in the state k+q emits a phonon within a plane wave-vector components q, where k is the 2D wave-vector, for the carrier with energy *E*. The second term arises, when a carrier in the same state k, absorbs a phonon with q. Here, *e* is the electronic charge, $\omega_0$ thee LO phonon angular frequency, $q_z$ the components of the phonon wave-vector Q, perpendicular to the interface, and $f_0(E)$ the Fermi-Dirac distribution function, at the electron temperature $T_c$, $\varepsilon_o$ denotes the free-space permittivity, $K_s$ and $K_\infty$ the static and optical dielectric constants of the static and optical dielectric constants of the material, respectively and $\sin\theta_0 = \{1-(k_{min}/k)^2\}^{1/2}$, where $k_{min} = [(m^*\omega_0)/hq-q/2]$.

The term $F^2(q_z L_z)$ is defined by:
$F^2(q_z L_z) = [\{\pi^2 \sin(q_z L_z/2)\}/\{\pi^2-(q_z L_z)^2\}(q_z L_z/2)]^2$
and
$(\delta N_Q/\delta t)_{decay} = [-(N_Q - \overline{N_Q})/\tau]$  …(9)

where $\overline{N_Q}$ is the thermal equilibrium value of the phonon distribution and $\tau$ the phonon life-time. Inserting Eqs (2-5) into Eq. (8) and the energy and momentum balance equations for the carriers terms up to first order are retained in the coefficient of $\sin\omega t$ and $\cos\omega t$ *on both* sides of the resulting equations are then equated to obtain the following set of equations:

$N_{Q0} = Q(T_0) N_Q + \tau Q_2(T_0)$ … (10)
$p_0 F_0 = b_1(T_0, N_{Q0})$ … (11)
$F_0/p_0 = b_2(T_0, N_{Q0})$ … (12)
$N_{Q1}[a_{3(T0)} + j\omega\tau] = \tau a_4(T_0)$ … (13)
$b_2(T_0, N_{Q0})(p_1/F_1) - b_3(T_0, N_{Q0})(T_1/F_1) + g_1(T_0, N_{Qr}) - jh_1(T_0, N_{Qi})(p_1/F_1) = 1$ … (14)
and
$b_4((T_0, N_{Q0}) + j\omega)(p_1/eF_1) + b_2(T_0, N_{Q0})(T_1/F_1) - g_1(T_0, N_{Qr}) - jh_1(T_0, N_{Qi})(p_1/F_1) = 1.$

… (15)

The functions, $a_i(i = 1-4)$, $b_i(i = 1-5)$, $g_i(i = 1-2)$, $h_i(i = 1-2)$ are determined by scattering mechanisms and the system parameters. Here, the detailed expressions for those functions have, been omitted for brevity, but they can be obtained in a straightforward way, following the procedure mentioned above. If non-equilibrium phonons are absent, then one can have:

$N_{Q0} = N_{Qi} = 0$ and $N_{Q0} = \overline{N_Q}$

For a given bias field $F_0$, the quantities $T_0$, $p_0$, $N_{Q0}$ and $N_{Qi}$ are found, by solving Esq. (11)-(14). The ac mobility $(p_1/m^*F_1)$ is then obtained from Esq. (15) and (16). The frequency at which ac mobility dopes to 0.707 of its low frequency value is called the 3-dB cut-off frequency, which is taken here, as a high frequency limit to the satisfactory response of the system.

## 3. RESULT AND DISCUSSION

Calculations are done numerically for a semiconductor quantum structures also called nanostructures of GaN with the material parameters of ref [3]. LO phonon life time is taken to be 7 pico seconds [6]. The numerical results so calculated in the present work, predict high frequency performance of GaN nanostructures and they can be used to analyze the experimental data, when they will appear in the literature. The ac mobility is found to decrease with increasing frequency beyond 80 GHz while drift velocity lags behind the applied field above around 220 GHz.

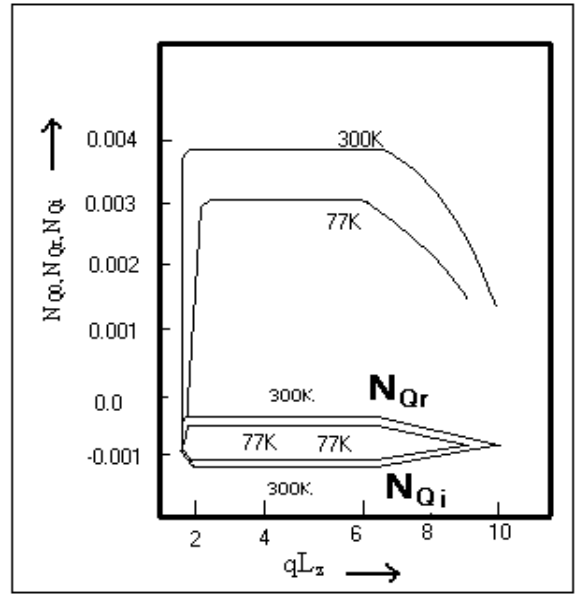

**Fig1: Variation of $N_{Q0}$, $N_{Qr}$, $N_{Qi}$ with $qL_z$ at 77K for LO phonons for $n_{2D}=10^{16} m^{-2}$, and $q_z=0$**

The $f_{3-dB}$ cut-off frequency at which the ac mobility drops to 0.707 of its low frequency value is found to decrease with increasing $L_Z$ and carrier concentration. The weakening of the scattering for a wider channel width, combined scattering mechanisms and the increasing degeneracy of the carriers at higher concentration account for such behaviour .however, its always higher when degradation of phonon distribution is incorporated .This behaviour is attributed to the slowing down of the carrier energy and momentum loss rates in the presence of non –equilibrium phonons [2].

Thus we find that the degradation of phonon distribution enhances significantly the 3-dB cut-off frequency thereby makes the high frequency response falter reflecting that high frequency response is better if effects of non-equilibrium phonon distribution are included in the calculations.

Fig.1 depicts the variation of $N_Q$, $N_{Qr}$ and $N_{Qi}$ versus $qL_z$ with $q_z = 0$, where $q$ and $q_z$ are, respectively, components of the phonon wave vector Q in-plane and perpendicular to the interface. The quantities are fairly constant in the mid-region, where as they are small for low as well as high values of $qL_z$. Less electron –LO phonon interaction at low $qL_z$ and small scattering probably for large-q LO phonon account for such variation. Infact, their values are controlled by the system parameters.

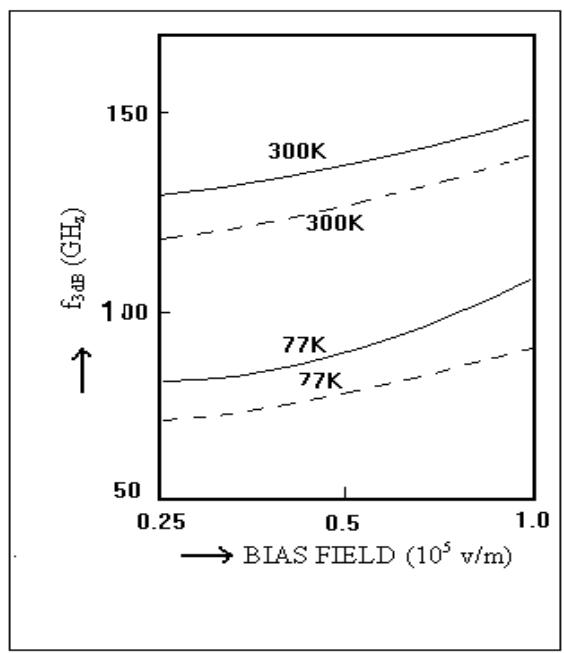

**Fig: Plot of $f_{3dB}$ with dc bias field for lattice temperature of 77K and 300K for $n_{2D}=10^{16}m^{-2}$ and $n_{bi}= 6 \times 10^{21}m^{-3}$. The solid curves include the hot phonon effects and the dashed curves donot.**

The biasing field dependence of 3-dB cut-off frequency are shown in the fig.2 $f_{3-dB}$ is found to increase with the bias field for both 77K and 300K and is higher at 300K than at 77K. The dashed curves give the result obtained without incorporating the hot phonon effects. It is evident from the figure that the inclusion of hot phonons increases significantly the 3-dB cut-off frequency.

In conclusion we find that the inclusion of non equilibrium LO phonons in the calculations enhances the 3-dB cut-off frequency significantly and makes high frequency response flatter reflecting that high frequency response is better if we include hot phonon effects in the calculations.

## 5. ACKNOWLEDGEMENTS

Subir Kumar Sarkar thankfully acknowledges the financial support of DRDO, Govt. of India.

## 4. CONCLUSION